\documentclass[sigconf]{acmart}

\copyrightyear{2022} 
\acmYear{2022} 
\setcopyright{rightsretained} 

\acmConference[CIKM '22]{Proceedings of the 31st ACM International Conference on Information and Knowledge Management}{October 17--21, 2022}{Atlanta, GA, USA}
\acmBooktitle{Proceedings of the 31st ACM Int'l Conference on Information and Knowledge Management (CIKM '22), Oct. 17--21, 2022, Atlanta, GA, USA}
\acmDOI{10.1145/3511808.3557700}
\acmISBN{978-1-4503-9236-5/22/10}

\usepackage{etoolbox}
\makeatletter
\patchcmd{\maketitle}{\@copyrightpermission}{
   \begin{minipage}{0.3\columnwidth}
     \href{https://creativecommons.org/licenses/by/4.0/}{\includegraphics[width=0.90\textwidth]{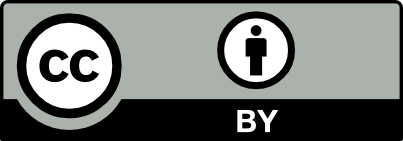}}
   \end{minipage}\hfill
   \begin{minipage}{0.7\columnwidth}
     \href{https://creativecommons.org/licenses/by/4.0/}{This work is licensed under a Creative Commons Attribution International 4.0 License.}
   \end{minipage}
  
   \vspace{5pt}
}{}{}

\makeatother

\usepackage{multirow}
\usepackage{booktabs}
\usepackage{bm}
\usepackage{amsfonts}
\usepackage{subcaption}
\usepackage{xspace}
\usepackage{enumitem}
\newcommand{\mname}{\texttt{SERF}\xspace}

\begin{document}

\title{\mname: Interpretable Sleep Staging using Embeddings, Rules, and Features}


\author{Irfan Al-Hussaini}
\affiliation{%
  \institution{Georgia Institute of Technology}
  \streetaddress{North Ave NW}
  \city{Atlanta}
  \state{Georgia}
  \country{USA}
  \postcode{303321}
}
\email{alhussaini.irfan@gatech.edu}
\author{Cassie S. Mitchell}
\affiliation{%
  \institution{Georgia Institute of Technology}
  \streetaddress{313 Ferst Drive}
  \city{Atlanta}
  \state{Georgia}
  \country{USA}
  \postcode{30332}
}
\email{cassie.mitchell@bme.gatech.edu}

\renewcommand{\shortauthors}{Irfan Al-Hussaini \& Cassie S. Mitchell}

\begin{abstract}
The accuracy of recent deep learning based clinical decision support systems is promising. However, lack of model interpretability remains an obstacle to widespread adoption of artificial intelligence in healthcare. Using sleep as a case study, we propose a generalizable method to combine clinical interpretability with high accuracy derived from black-box deep learning.

Clinician-determined sleep stages from polysomnogram (PSG) remain the gold standard for evaluating sleep quality. However, PSG manual annotation by experts is expensive and time-prohibitive. We propose \mname, \textit{interpretable Sleep staging using Embeddings, Rules, and Features} to read PSG. \mname provides interpretation of classified sleep stages through meaningful features derived from the AASM Manual for the Scoring of Sleep and Associated Events.

In \mname, the embeddings obtained from a hybrid of convolutional and recurrent neural networks are transposed to the interpretable feature space. These representative interpretable features are used to train simple models like a shallow decision tree for classification. Model results are validated on two publicly available datasets. \mname surpasses the current state-of-the-art for interpretable sleep staging by 2\%. Using Gradient Boosted Trees as the classifier, SERF obtains 0.766 $\kappa$ and 0.870 AUC-ROC, within 2\% of the current state-of-the-art black-box models.
\end{abstract}






\maketitle

\section{Introduction}
\label{sec:intro}
The prevalence of electronic health records has led to an abundance of patient health data \cite{adler2017hitech, jensen2012mining}. Meanwhile, recent advances in deep learning have shown great promise in utilizing these data for accurate clinical decision support systems. However, the lack of interpretability remains an obstacle to widespread adoption for a high stakes application like healthcare \cite{zitnik, elshawi2019interpretability, carvalho2019machine, holzinger2019causability, wiens2019no}. On the other hand, simple linear models are not accurate enough to be used by clinicians for decision-making \cite{london2019artificial}. Can a clinical decision support system be developed that is interpretable and accurate?

In this paper, we focus on sleep staging. Sleep stages annotated by clinicians from polysomnograms (PSG) remain the gold standard for evaluating sleep quality and diagnosing sleep disorders. 

Sleep disorders affect 50–70 million US adults and 150 million in developing countries worldwide \cite{guglietta2015drug}.
Sleep staging is the most important precursor to sleep disorder diagnoses such as insomnia, narcolepsy, or sleep apnea \cite{RN50}.
However, the clinician-determined sleep staging, which is the gold standard, is labor-intensive and expensive \cite{guillot2020dreem}. 
Neurologists visually analyze multi-channel PSG signals and provide empirical scores of sleep stages, including wake, rapid eye movement (REM), and the non-REM stages N1, N2, and N3, following guidelines stated in the American Academy of Sleep Medicine (AASM) Manual for the Scoring of Sleep and Associated Events \cite{RN4}. Such a visual task is cumbersome and takes several hours for one sleep expert to annotate a patient’s PSG signals from a single night \cite{ZHANG2022100371}.

Automated algorithms for sleep staging alleviate these limitations. Deep learning methods have successfully automated annotation of sleep stages by using convolutional neural networks (CNN) \cite{RN49, RN57, RN9, yang2021single}, recurrent neural networks \cite{RN14, phan2019seqsleepnet}, recurrent convolutional neural networks \cite{RN6, RN62}, deep belief nets \cite{RN28}, autoencoders \cite{RN63}, attention \cite{qu2020residual, phan2022sleeptransformer, li2022attention},  and graph convolutional neural networks \cite{li2022attention, jia2020graphsleepnet}. 
Although deep learning models can produce accurate sleep staging classification, they are often treated as black-box models that lack interpretability \cite{RN34}. Lack of interpretability limits the adoption of the deep learning models in practice because clinicians must understand the reason behind each classification to avoid data noise and unexpected bias \cite{zitnik, mlhc_int}. Furthermore, current clinical practice at sleep labs relies on the AASM sleep scoring manual \cite{RN4}, which is interpretable for clinical experts but lacks precise definitions needed for a robust computational model \cite{al2019sleeper}. 


Thus, an automated model for sleep staging should ideally be as clinically interpretable as the sleep scoring manual and as accurate as the black-box neural network models. 
To this end, we propose \mname, \textit{interpretable Sleep staging using Embeddings, Rules, and Features}, which combines clinical interpretability with the accuracy derived from a deep learning model. It provides clinically meaningful explanations, derived from the AASM manual \cite{RN4}, with each sleep stage prediction. \mname surpasses the interpretable sleep staging baseline \cite{al2019sleeper} and performs within 2\% of state-of-the-art black-box deep learning models \cite{RN57}.


\section{Data}
\label{sec:data}
The following two publicly available datasets, summarized in Table \ref{tab:dataset}, are used to evaluate the performance of SERF:
\begin{itemize}[leftmargin=*]
    \item PhysioNet EDFX \cite{physionet1, physionet2}: The PhysioNet EDFX database contains 197 whole-night PSG sleep recordings. The corresponding hypnograms were scored by technicians according to the Rechtschaffen and Kales guidelines \cite{rk}. 
    Sleep Stages N3 and N4 were combined to adhere to AASM standards. 153 subjects were healthy Caucasians without any sleep-related medication. 44 subjects had mild difficulty falling asleep. It contains two Electroencephalography (EEG) channels, one Electrooculography (EOG) channel, and one Electromyography (EMG) channel sampled at 100 Hz.
    \item ISRUC \cite{RN23}:  The ISRUC dataset contains PSG recordings of 100 subjects with evidence of having sleep disorders. The data was collected from 55 male and 45 female subjects, whose ages range from 20 to 85 years old, with an average age of 51. The corresponding hypnograms were manually scored by technicians according to the AASM manual \cite{RN4}. It includes six EEG channels (F3, F4, C3, C4, O1, and O2), two EOG channels (E1 and E2), and a single EMG channel sampled at 200 Hz.

\end{itemize}
\begin{table}[tb]
  \begin{minipage}{\linewidth}
    \centering
\caption[for LOF]{Datasets}
\resizebox{\linewidth}{!}{%
\begin{tabular}{lcccc}
\hline
                        & \textbf{\begin{tabular}[c]{@{}c@{}}Number of \\ Subjects\end{tabular}} & \textbf{\begin{tabular}[c]{@{}c@{}}Sampling \\ Frequency (Hz)\end{tabular}} & \textbf{\begin{tabular}[c]{@{}c@{}}Number of \\ Channels\end{tabular}} & \textbf{\begin{tabular}[c]{@{}c@{}}Annotation \\ Schema\end{tabular}} \\ \hline
\textbf{ISRUC} \cite{RN23}          & 100                                                                    & 200                                                                         & 9                                                                      & AASM \cite{RN4} \\ 
\textbf{PhysioNet-EDFX} \cite{physionet1, physionet2} & 197                                                                    & 100                                                                         & 4                                                                      & R\&K \cite{rk, rk2}                       \\ \hline
\end{tabular}}
\label{tab:dataset}
\end{minipage}
\end{table}
\section{Method}
\label{sec:method}


\begin{figure}[htb]
    \centering
    \includegraphics[ width=\linewidth, keepaspectratio]{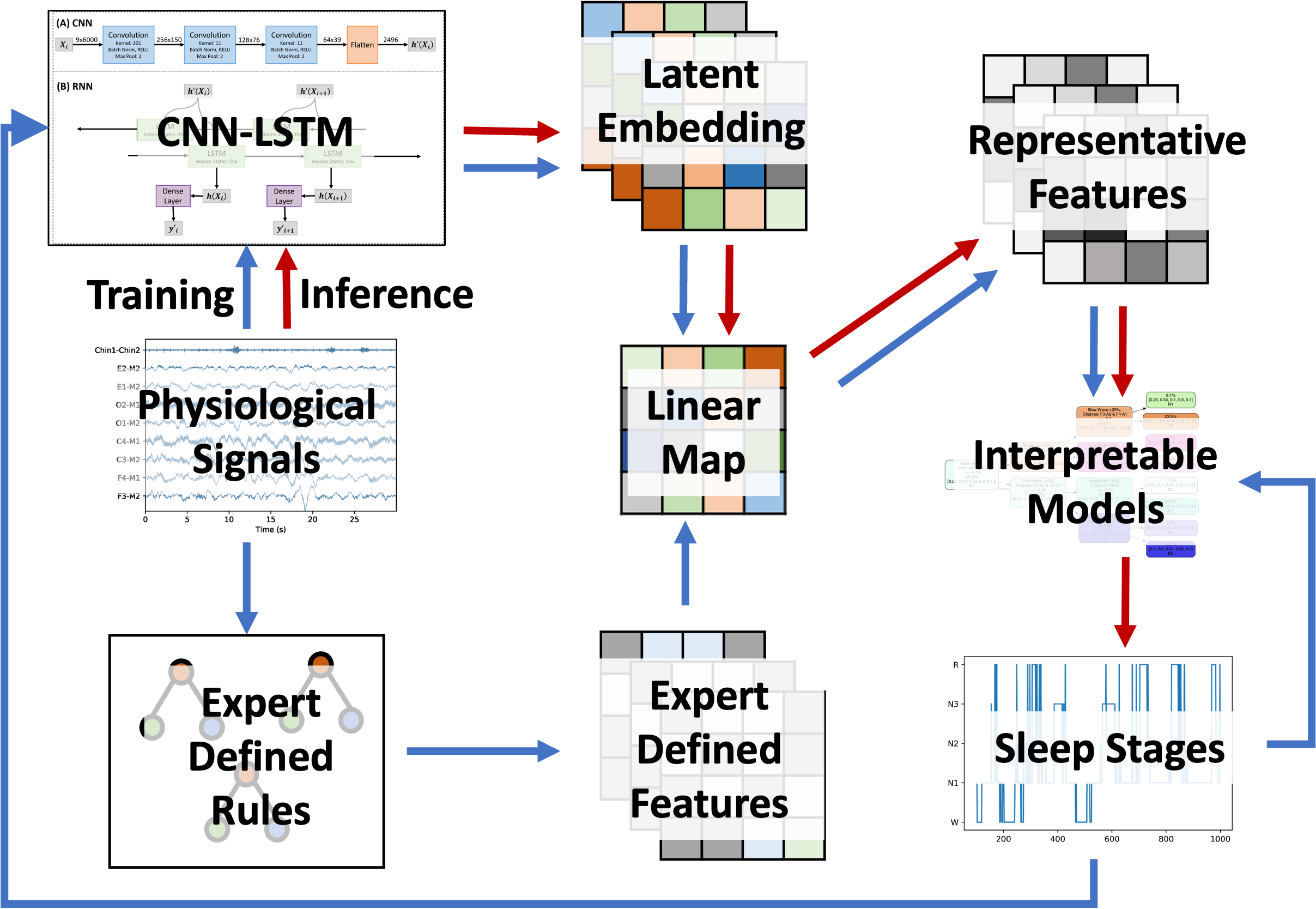}
\caption{\mname Framework}
\label{fig:model}
\end{figure}

\mname predicts sleep stages using PSG data through an interpretable model derived from expert-defined features and embeddings from a deep neural network. The method is explained using the ISRUC dataset. Table \ref{tab:dataset} and Section \ref{sec:data} can be used to find the corresponding metrics for PhysioNet-EDFX.

The input data are multi-channel PSG signals. It consists of multivariate continuous time-series data, $\mathcal{X}$, partitioned into $N$ segments called epochs, denoted as $\mathcal{X} = \{\bm{X}_1, \cdots, \bm{X}_N\}$. Each epoch $\bm{X}_n \in \mathbb{R}^{9\times6,000}$ is 30 seconds long and contains 9 physiological signals from 9 channels. The sampling frequency is 200Hz.
Each epoch, $\bm{X}_n$, also has an associated sleep stage label $y_n \in \{\text{Wake, REM, N1, N2, N3}\}$ provided by a clinical expert. The goal is to predict a sequence of sleep stages, $\bm{S}=\{s_1,\; \ldots\; s_N\}$ based on $\mathcal{X}$ so that they mimic the labels provided by human experts, $\bm{Y} = \{y_1,\; \ldots\; y_N\}$. In \mname, these predictions also contain meaningful explanations derived from expert-defined features. The \mname framework shown in Figure \ref{fig:model}  comprises the following steps:
\subsection{Latent Embedding}
The multivariate PSG signals are embedded using CNN-LSTM to capture translation invariant and complex patterns. The CNN is composed of 3 convolutional layers. Each convolutional layer is followed by batch normalization, ReLU activation, and max pooling. Using a kernel size of 201, the convolutions in the first layer extract features based on 1-second segments. Subsequent layers have a kernel size of 11. The output channels of the three convolution layers are 256, 128, and 64. The output of the final convolutional layer is flattened and fed into a single layer of bi-directional Long Short-Term Memory (LSTM) cells with 256 hidden states to capture temporal relations between epochs. This results in 512 hidden states for each epoch, $\bm{X}_i$ and represents the latent embedding used in subsequent steps:
$$\bm{h}(\bm{X}_i )\in \mathbb{R}^{512}$$
A single fully connected layer with softmax activation is then used to predict the five sleep stages:
$$\bm{z}_i=\bm{W}^T h(\bm{X}_i )+\bm{b}$$
$$\bm{s}_i=softmax(\bm{z}_i)$$
where $\mathbf{W}\in\mathbb{R}^{512\times5}$ is the weight matrix, $\mathbf{b}\in\mathbb{R}^5$ is the bias vector, and $\bm{s}_i$ is the estimated probabilities of all 5 sleep stages at epoch i. Cross-entropy loss is used to train the model:
$$L(\bm{y}_i, \bm{s}_i )=-\sum_j^5\bm{y}_i [j]  log(\bm{s}_i [j])$$
where $L\left(\bm{y}_i,\;\bm{s}_i\right)$ is the estimated cross-entropy loss for epoch i between human labels $\bm{y}_i$ and the predicted probabilities $\bm{s}_i$. After training on sleep stage prediction, the latent embedding $\bm{h}\left(\mathcal{X}\right) \in \mathbb{R}^{N \times 512}$ is obtained from the hidden states of the LSTM.

\subsection{Expert Defined Features}
Concurrently, each epoch is encoded into a feature vector. Expert suggestions are incorporated to supplement the technical guidelines in the AASM manual \cite{RN4}. Using this rule augmentation procedure, a set of $M'$ features are extracted, $\bm{F'}\left(\bm{X_n}\right) = \left[f'_1\left(\bm{X}_n\right), \ldots f'_{M'}\left(\bm{X}_{n}\right)\right]$, where element $f'_j\left(\bm{X}_n\right)$ is the function generating feature j for epoch $\bm{X}_n$. 

\noindent These meaningful features are described below:
\begin{itemize}[leftmargin=*]
	\item Sleep spindles are bursts of oscillatory signals originating from the thalamus and depicted in EEG \cite{RN13}. It is a discriminatory feature of N2. 
	\item Slow-wave sleep is prominent in N3 and is marked by low-frequency and high-amplitude delta activity \cite{whitmore2022targeted}. 
	\item Delta (0.5–4Hz) waves feature in N3, Theta (4–8 Hz) in N1, Alpha (8–12 Hz), and Beta ($>$12 Hz) help to differentiate between Wake and N1 \cite{keenan2011monitoring}. EMG bands determine the muscle tone used to distinguish between REM and Wake \cite{naylor2011simultaneous}. Each band's Power Spectral Density (PSD) is calculated using a multi-taper spectrogram.
	\item Amplitude is vital in finding K Complexes, Chin EMG amplitude, and Low Amplitude Mixed Frequency and thus used to differentiate Wake, REM, N1, and N2. 
	\item Mean, Variance, Kurtosis, and Skew are used to capture the distribution of values in the channel. It can help detect outlier values that highlight signatures such as K-Complexes.
\end{itemize}
The importance of expert-defined features is analyzed using the ANOVA test to select the top 90\% of the most discriminative features, $\bm{F}$. It reduces the number of features from $M'$ to $M$, where $\bm{F}\subset \bm{F}'$. 
These $M$ features from $N$ epochs leads to our feature matrix $\bm{F}\left(\mathcal{X}\right)\in\mathbb{R}^{N\times M}$, which forms the basis of the interpretation module of our framework where an element $f_j\left(\bm{X}_i\right)$ is the value of feature j at the epoch $\bm{X}_i$. Number of features in ISRUC dataset: $M'=87$ and $M=78$, and in Physionet dataset:  $M'=38$ and $M=34$ due to fewer channels.
\subsection{Linear Map}
A linear map, $\bm{T}$, combines the inputs encoded by features and latent embeddings. After the feature matrix, $\bm{F}\left(\mathcal{X}\right)$, and latent CNN-LSTM embedding matrix, $\bm{h}\left(\mathcal{X}\right)$, are generated, a linear transformation $\bm{T}$ is learned, which maps the features into the latent space defined by the embeddings. The linear transformation matrix, $\bm{T}$, is learned using ridge regression: 

$$\min_T\left||\bm{h}\left(\mathcal{X}\right)-\bm{F}\left(\mathcal{X}\right)\bm{T}|\right|_2^2 + ||\bm{T}||_2^2$$

\subsection{Representative Features}
The linear map, $\bm{T}$, is used to generate a representative feature, $\bm{s}_j$, for each epoch using the embedding vectors $\bm{h}\left(\bm{X}_n\right)$. These representative features collectively form the similarity matrix $\bm{S}$ and is used to train an interpretable classifier like a shallow decision tree.

Given an epoch,  $\bm{X}_j$, the CNN-LSTM embedding module is used to obtain the embedding, $\bm{h}\left(\bm{X}_j\right)$. A representative feature similarity matrix, $\bm{S}$, is then generated using the linear map, T:
$$\bm{S}=\bm{h}(\bm{X}_j)\bm{T}^T$$

This representative feature similarity matrix is used as input to simple classifiers such as a shallow decision tree. When a new PSG is provided, $\bm{X}$, the embedding vector $\bm{h}(\bm{X})$ is first generated using the CNN-LSTM network followed by the representative feature similarity matrix $\bm{S}=\bm{h}(\bm{X}_i)\bm{T}^T$. These representative features are used as input to simple classifiers and form the basis for model interpretability.  

\section{Experiments}
\label{sec:experiments}
\textbf{Implementation Details}: \mname was built using PyTorch 1.0 \cite{NEURIPS2019_bdbca288}, scikit-learn \cite{RN44}, and XGBoost \cite{xgboost}. A batch size of 1000 samples from 1 PSG is used. Each model is trained for 20 epochs with a learning rate of $10^{-4}$ using ADAM as the optimization method.
The data is randomly split by subjects into a training and test set in a 9:1 ratio with the same seed for each experiment. For each dataset, the training set is used to fix model parameters, and the test set is used to obtain performance metrics. The same model hyperparameters and feature extraction schema are used to prevent overfitting and ensure consistent performance across different datasets.
\\\textbf{Baselines:}
\begin{itemize}[leftmargin=*]
	\item Convolutional Neural Network with a stacked bi-directional LSTM layer (CNN-LSTM): the black-box model used in obtaining the signal embeddings. 
	\item 1D-Convolutional Neural Network (1D-CNN): a black-box model proposed in \cite{al2019sleeper}.
	\item Expert Feature Matrix, $\bm{F}\left(\mathcal{X}\right)$, as input to simple classifiers.
	\item SLEEPER \cite{al2019sleeper}: an interpretable sleep staging algorithm based on prototypes.
	\item U-Time \cite{RN57}: state-of-the-art black-box deep learning model which adapts the U-Net architecture for sleep staging.
\end{itemize}
\textbf{Metrics:}
\begin{itemize}[leftmargin=*]
    \item Accuracy $=\frac{\left|\mathcal{Y}\cap\mathcal{Y}^\prime\right|}{N}$
    \item Sensitivity, $S^{\left(k\right)}=\frac{\left|\mathcal{Y}^{\left(k\right)}\cap\mathcal{Y}^{\prime\left(k\right)}\right|}{\left|\mathcal{Y}^{\prime\left(k\right)}\right|}$\item Precision, $P^{\left(k\right)}=\frac{\left|\mathcal{Y}^{\left(k\right)}\cap\mathcal{Y}^{\prime\left(k\right)}\right|}{\left|\mathcal{Y}^{\left(k\right)}\right|}$ 
    \item F1 score $=\frac{2\ \ast\ P\ \ast\ S}{P\ +\ S\ }$
    \item Cohen's $\kappa=\frac{Acc-p_e}{1-p_e}$, where $p_e=\frac{1}{N^2}\sum_{k}^{5}\left|\mathcal{Y}^{\left(k\right)}\right|\left|\mathcal{Y}^{\prime\left(k\right)}\right|$
\end{itemize}
Given expert annotations $\mathcal{Y}'$ and predicted stages $\mathcal{Y}$ of size $N$, $k=\{W,N1,N2,N3,R\}$ indicating the sleep stage, and $\left|{\mathcal{Y}^\prime}^{\left(k\right)}\right|$ $(\left|\mathcal{Y}^{\left(k\right)}\right|)$ is the number of human (algorithm) labels from sleep stage k.
\begin{table}[tb]
  \begin{minipage}{\linewidth}
    \centering
\caption[LOR]{Model Evaluation\footnote{XG: DART Gradient Boosted Trees, DT: Decision Tree, LR: Logistic Regression, GB: Gradient Boosted Trees}}
\resizebox{\linewidth}{!}{%
\begin{tabular}{l@{\qquad}cc@{\qquad}cc@{\qquad}cc@{\qquad}cc}
  \toprule
  \multirow{2}{*}{\raisebox{-\heavyrulewidth}{Model}} & \multicolumn{2}{c}{Accuracy $(\%)$} & \multicolumn{2}{c}{ROC-AUC $(\%)$} & \multicolumn{2}{c}{Cohen's $\kappa$} & \multicolumn{2}{c}{Macro F1}\\
  \cmidrule{2-9}
  
                             & EDFx          & ISRUC & EDFx         & ISRUC & EDFx      & ISRUC & EDFx      & ISRUC\\
\midrule
\mname-DT & 81.2 & 80.5 & 82.5 & 85.8 & 0.735 & 0.747 & 0.719 & 0.768 \\
\mname-XG & 82.3 & 81.9 & 84.4 & 87.0 & 0.753 & 0.766 & 0.753 & 0.789\\
\mname-GB &82.2 & 81.7 & 84.8 & 87.0 & 0.753 & 0.763 & 0.758 & 0.789 \\
\mname-LR & 82.9 & 79.5 & 85.0 & 85.3 & 0.762 & 0.733 & 0.759 & 0.773\\
Features-XG & 81.0 & 77.4 & 83.2 & 83.0 & 0.734 & 0.704 & 0.732 & 0.722 \\
Features-DT & 68.7 & 71.6 & 74.3 & 79.4 & 0.555 & 0.629 & 0.583 & 0.665 \\
SLEEPER-DT \cite{al2019sleeper} & 78.8 & 78.0  & 81.5 & 83.4 & 0.704 & 0.712 & 0.696 & 0.730 \\
SLEEPER-GB \cite{al2019sleeper} & 80.7 & 79.7  & 82.8 & 85.1 & 0.729 & 0.736 & 0.721 & 0.756 \\
U-Time \cite{RN57} & 86.2 & 84 & 88.3 & 88.8 & 0.810 & 0.793 & 0.811 & 0.816 \\
CNN-LSTM & 86.4 & 83.1 & 88.6 & 88.8 & 0.813 & 0.783 & 0.815 & 0.819\\
1D-CNN \cite{al2019sleeper} & 84.4 & 82.5 & 86.6 & 87.2 & 0.784 & 0.773 & 0784 &  0.789\\
  \bottomrule
\end{tabular}}
\label{tab:eval}
\end{minipage}
\end{table}
\begin{table}[htb]
  \begin{minipage}{\linewidth}
    \centering
\caption{Sensitivity across Sleep Stages}
\resizebox{\linewidth}{!}{%
\begin{tabular}{l@{\qquad}cc@{\qquad}cc@{\qquad}cc@{\qquad}cc@{\qquad}cc}
  \toprule
  \multirow{2}{*}{\raisebox{-\heavyrulewidth}{Model}} & \multicolumn{2}{c}{Wake} & \multicolumn{2}{c}{N1} & \multicolumn{2}{c}{N2} & \multicolumn{2}{c}{N3}& \multicolumn{2}{c}{R}\\
  \cmidrule{2-11}
  
                             & EDFx          & ISRUC & EDFx         & ISRUC & EDFx      & ISRUC & EDFx      & ISRUC & EDFx      & ISRUC\\
\midrule
\mname-DT & 0.897 & 0.889 & 0.283 & 0.440 & 0.849 & 0.798 & 0.803 & 0.855 & 0.763 & 0.859\\
\mname-XG & 0.892 & 0.894 & 0.386 & 0.481 & 0.863 & 0.809 & 0.823 & 0.889 & 0.802 & 0.870 \\
\mname-GB & 0.892 & 0.888 & 0.410 & 0.495 & 0.863 & 0.811 & 0.824 & 0.888 & 0.802 & 0.865\\
\mname-LR & 0.904 & 0.895 & 0.398 & 0.535 & 0.865 & 0.768 & 0.832 & 0.847 & 0.799 & 0.819 \\
Features-XG & 0.891 & 0.873 & 0.340 & 0.324 & 0.842 & 0.768 & 0.828 & 0.859 & 0.761 & 0.788\\
Features-DT & 0.757 & 0.821 & 0.110 & 0.250 & 0.742 & 0.717 & 0.692 & 0.789 & 0.612 & 0.749 \\
SLEEPER-DT \cite{al2019sleeper} & 0.902 & 0.893 & 0.256 & 0.329 & 0.840 & 0.764 & 0.819 & 0.865 & 0.661 & 0.799 \\
SLEEPER-GB \cite{al2019sleeper} & 0.911 & 0.902 & 0.313 & 0.397 & 0.848 & 0.787 & 0.835 & 0.876 & 0.700 & 0.816 \\
U-Time \cite{RN57} & 0.941 & 0.890 & 0.552 & 0.504 & 0.897 & 0.839 & 0.749 & 0.898 & 0.883 & 0.951 \\
CNN-LSTM & 0.936 & 0.871 & 0.507 & 0.548 & 0.905 & 0.781 & 0.838 & 0.937 & 0.859 & 0.956  \\
1D-CNN \cite{al2019sleeper} & 0.926 & 0.914 & 0.397 & 0.398 & 0.910 & 0.840 & 0.817 & 0.886 & 0.819 & 0.907 \\
  \bottomrule
\end{tabular}}
\label{tab:stages}
\end{minipage}
\end{table}
\\\textbf{Results:}
The results from experiments are compared in Table \ref{tab:eval} and \ref{tab:stages}.
\mname performs within 2\% of state-of-the-art black box models such as U-Time \cite{RN57} and far exceeds the performance of expert feature-based models. Table \ref{tab:stages} shows that for all sleep stages other than Wake, \mname surpassed SLEEPER \cite{al2019sleeper} and was comparable to black-box models,  U-Time \cite{RN57} and CNN-LSTM. N1 is a challenging sleep stage to identify where \mname was comparable to black-box models. These results show that \mname can generalize better than other interpretable methods \cite{al2019sleeper} and is comparable to black-box models \cite{RN57} in identifying all stages.

 \mname has the following improvements over the interpretable method, SLEEPER \cite{al2019sleeper}: (1) \mname surpasses SLEEPER by 2-3\% in all evaluation metrics. (2) \mname has a lower representative feature dimension by utilizing raw feature values instead of binary rules resulting in smaller matrices. (3) \mname learns a linear map using ridge regression with a lower dimension than the prototype matrix learned in SLEEPER using cosine similarity, thus resulting in a smaller model size. (4) The smaller feature size of \mname results in faster training and faster inference from the simple classifiers. (5) Meaningful feature value cutoffs are obtained at nodes of the decision tree, as seen in Figure \ref{fig:tree}, instead of just similarity indices.

The results also show the significance of different channels when building an interpretable model. The clinical sleep staging manual \cite{RN4} utilizes all the 9 channels in the ISRUC dataset. Since the Physionet EDFx dataset only contains 4 channels there are some features that cannot be extracted. As a result, \mname, SLEEPER \cite{al2019sleeper} and Feature models exhibit worse performance relative to black-box models for the EDFx dataset than ISRUC.


Figure 2 shows the significance of representative features for classifying each sleep stage based on \mname and gradient boosted trees. We focus on 2 of the top 7 features, which can be better attributed to guidelines in the sleep staging manual \cite{RN4}. The rest of the top 7 features are general meaningful features rather than distinctive signal traits embedded in epochs.
The 3rd feature, Spindle in the C3-A2 and C4-A1 contra-lateral channel pairs, is important in identifying REM stages. The manual states the absence of Spindles as a critical observation when annotating REM. The 7th most impactful feature, Slow Wave in the C3-A2 and C4-A1 contra-lateral channel pairs, contributes significantly to the distinction of N3 from N2. The most distinctive attribute clinicians look for in N3 is a slow wave.
\section{Interpretation}
\label{sec:interpretation}
\begin{figure}[htb]
 \begin{minipage}{\linewidth}
    \centering
    \includegraphics[ width=\linewidth, keepaspectratio]{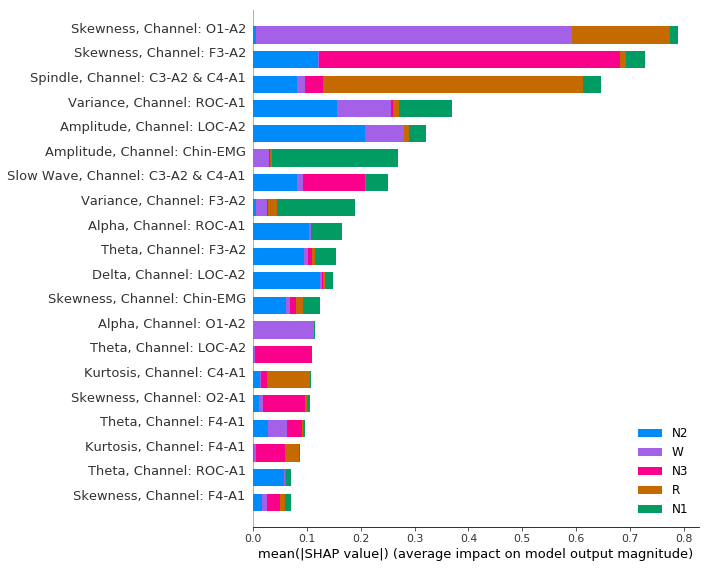}
    \caption[tmp LOF]{\mname-XG feature importance SHAP values (ISRUC)} 
    \label{fig:features}
\end{minipage}
\end{figure}

\begin{figure}[htb]
 \begin{minipage}{\linewidth}
    \centering
    \includegraphics[ width=\linewidth, keepaspectratio]{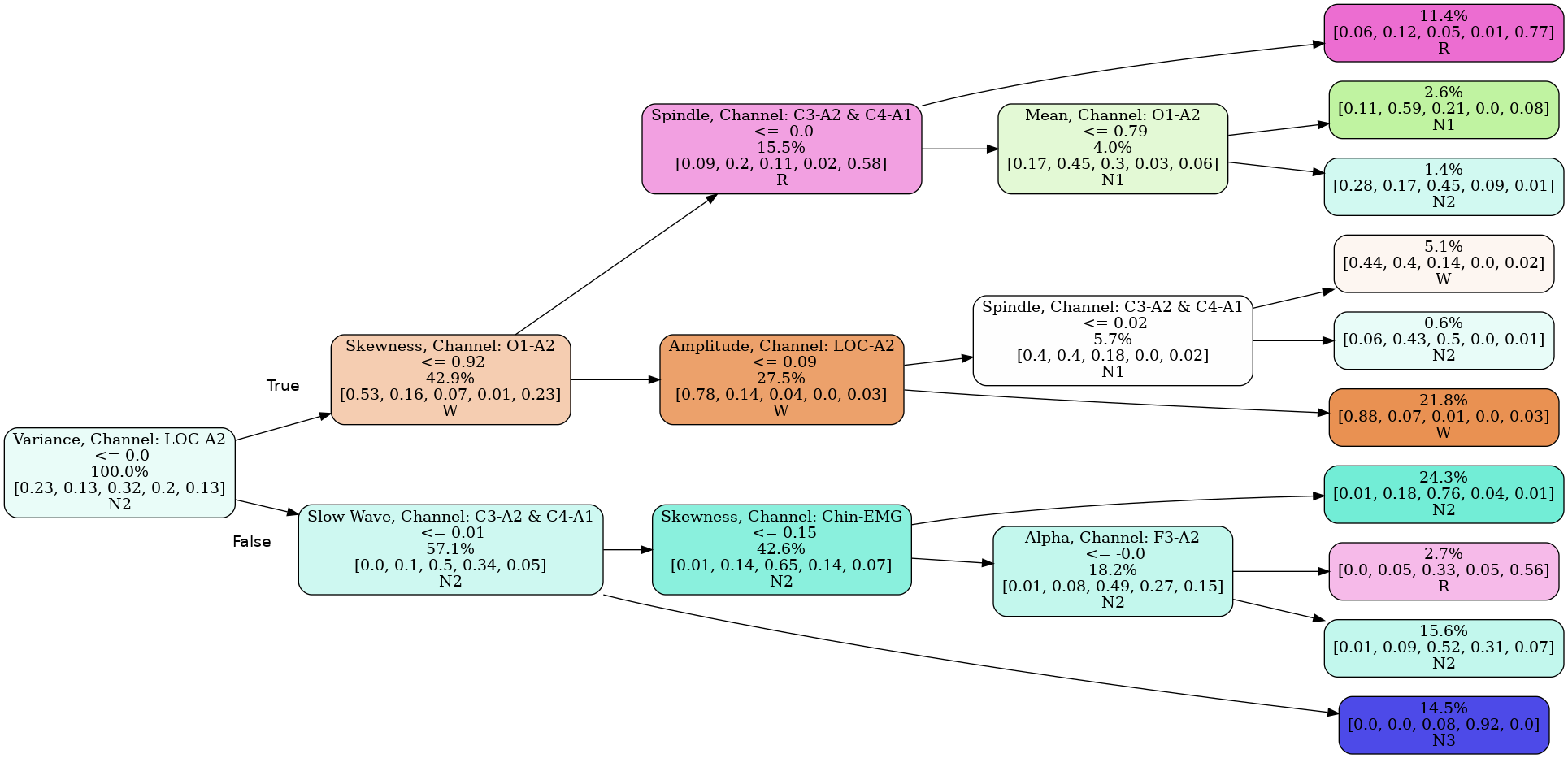}
    \caption[tmp LOF]{\mname-Decision Tree (ISRUC)
    } 
    \label{fig:tree}
\end{minipage}
\end{figure}

Figure 3 shows a decision tree of depth 4, based on \mname. 
The left-most node denotes the root of the tree. 
The color indicates the most frequent sleep stage at a node, and the intensity is proportional to its purity. 
The five rows of each node contain the following: (1) the feature and the channels used, (2) the feature cutoff value, (3) the percentage of data passing through, (4) the ratio of each sleep stage in the following order: [Wake, N1, N2, N3, REM], (5) the most frequent sleep stage, in other words if classification is performed at that node, this label is assigned. 
Analyzing the resulting decision tree reveals some promising aspects of \mname. According to the sleep staging guidelines for human annotators \cite{RN4}, N3 is distinguished by the occurrence of slow waves, one of the underlying features of \mname. The bottom leaf node is partitioned using Slow Waves $>= 0.01$ in the C3-A2 \& C4-A1 channel pair. 92\% of this leaf node contains N3, while only 34\% of the previous node contains it.

\section{Conclusion}
We propose a method, \mname, to provide accurate and interpretable clinical decision support and demonstrate it on automated sleep stage prediction. In order to achieve this goal, \mname combines embeddings from a deep neural network with clinically meaningful features. \mname achieves high performance metrics, comparable to state-of-the-art deep learning baselines. Moreover, the \mname expert feature module incorporates standard AASM guidelines to ensure the model enables transparent clinical interpretability, as illustrated using two qualitative case studies. 
\section*{Acknowledgments}

This research was funded by NSF 1944247, NIH U19-AG056169, GT McCamish Award to C.M.

\bibliographystyle{ACM-Reference-Format}
\balance
\bibliography{refs}
\end{document}